\documentclass[final,3p,times]{elsarticle}

\usepackage{lineno,hyperref}
\usepackage{amsthm,amsmath,amssymb,eucal}
\usepackage{caption}
\usepackage{subcaption}
\usepackage{mathdots}

\modulolinenumbers[5]

\journal{Physics Letters A}

\newcommand{\R}{\mathbb{R}}
\newcommand{\C}{\mathbb{C}}
\newcommand{\D}{\mathrm{d}}
\newcommand{\Z}{\mathbb{Z}}
\newcommand{\e}{\mathrm{e}}
\newcommand{\HH}{\mathcal{H}}
\newcommand{\OO}{\mathcal{O}}

\newdefinition{remark}{Remark}

\begin{document}

\begin{frontmatter}

\title{Quantum graphs: self-adjoint, and yet exhibiting a nontrivial $\mathcal{PT}$-symmetry}





\author[label1,label2]
{Pavel Exner}
\ead{exner@ujf.cas.cz}
\author[label2]
{Milo\v{s} Tater}
\ead{tater@ujf.cas.cz}
\address[label1]{Doppler Institute for Mathematical Physics and Applied Mathematics, Czech Technical University,
B\v rehov{\'a} 7, 11519 Prague, Czechia}
\address[label2]{Department of Theoretical Physics, Nuclear Physics Institute, Czech Academy of Sciences, 25068 \v{R}e\v{z} near Prague, Czechia}

\begin{abstract}
We demonstrate that a quantum graph exhibits a $\mathcal{PT}$-symmetry provided the coefficients in the condition describing the wave function matching at the vertices are circulant matrices; this symmetry is nontrivial if they are not invariant with respect to transposition. We also illustrate how the transport properties of such graphs are significantly influenced by the presence or absence of the non-Robin component of the coupling.
\end{abstract}


\end{frontmatter}

\linenumbers

Having two research areas, each based on a strong concept, one is naturally inspired to look for connecting links between them. This applies, in particular, to the concepts of quantum graphs \cite{BK} and $\mathcal{PT}$-symmetry \cite{CY}. Neglecting the prehistory of the former associated with the idea of Linus Pauling \cite{Pa}, they came to life roughly at the same time and experienced three decades of intense development. Quantum graphs were rediscovered as a versatile model of nanostructurs, but proved themselves a useful tool to investigate fundamental properties of quantum systems; as an example one can mention their role in analysis of quantum chaos \cite{KS}. $\mathcal{PT}$-symmetry, on the other hand, started from the observation that Schr\"odinger operators with complex potentials can have a real spectrum \cite{BB}, and while the importance of this fact for quantum mechanics remains a matter of dispute for those who are not $\mathcal{PT}$-proselytes, the idea found a number of applications in various areas.

Speaking of $\mathcal{PT}$-symmetry the focus was understandably on the nontrivial situations when neither parity nor the time-reversal invariance were preserved but their composition was. This motivated the use of complex potentials having in mind that Schr\"odinger operators with real ones are naturally time-reversal invariant. Sometimes, of course, a real potential emerges in this context, as is the case of the negative quartic one, but even then the ways in which it is treated may be very different \cite{Be, ZK}.

Various ways of introducing $\mathcal{PT}$-symmetry to quantum graphs have been proposed, all of them having in common that they go beyond the class of self-adjoint Hamiltonians. Mostly the authors extended the class of boundary conditions determining the vertex coupling, either globally \cite{AKU, HKS} or in a particular subset of vertices. For instance, in \cite{KMG} the coupling is supposed to be Kirchhoff in the internal vertices while at the endpoints of the `loose edges' complex Robin boundary conditions are imposed, in fact generalizing the very simple and elegant example from \cite{KBZ} in which the graph was just a line segment; we note that in these situations the Robin conditions have a natural interpretation as a source or drain. Another approach is to begin with a modification of the scalar product in the state Hilbert space \cite{MSY}; it is known that operator facilitating the appropriate similarity transformation may or may not exist \cite{KLZ}.

The main observation in this letter is that self-adjoint quantum \emph{may be} $\mathcal{PT}$-symmetric and parity violating at the same time, even if (a) there is no potential, that is, the only interaction comes from wave function matching at the graph vertices, and (b) the coefficients describing the latter are \emph{real}. To be more specific, we are going to demonstrate such a symmetry -- nontrivial whenever the vertex coupling violates the time-reversal invariance -- applies to graphs in which those coefficients are \emph{circulant matrices}. In addition, we are going to provide some insights concerning spectral and transport properties of quantum graph vertices underlying the role played by the balance between the three component of the coupling, the Dirichlet, Neumann, and Robin one \cite[Thm.~1.4.4]{BK}. We will also give an example how an arbitrarily small change of coupling parameters can completely change transport properties of a lattice in terms of the probability introduced in \cite{BaB}.

As usual the state Hilbert of a quantum graph is the direct sum, $\HH=\sum_j^\oplus L^2(e_j)$, over the graph edges and the Hamiltonian acts on the $j$th edge as $\psi_j\mapsto -\psi_j''$. Without loss of generality we may discuss a star graph, $N$ halflines meeting at a single vertex, because (a) the vertex coupling is local, and (b) more complicated graphs can be replaced by a graph with one vertex only, joining a certain number of halflines and loops, and the boundary conditions with matrix coefficients having a block structure reflecting the topology of the original graph. We use the symbols $\Psi=(\psi_1(0),\ldots,\psi_n(0))^T$ and $\Psi'=(\psi_1'(0),\ldots,\psi_n'(0))^T$, respectively, for the vectors of boundary values of the wave function components and their derivatives, conventionally taken in the outward direction, at the vertex. The boundary conditions $A\Psi+B\psi'=0$ ensure the self-adjointness provided the matrix $(A|B)$ has maximum rank and $B^*A$ is Hermitean \cite{KoS}; alternatively we can write them in the unique form,
\begin{equation}\label{bc}
(U-I)\Psi+i\ell(U+I)\Psi'=0,
\end{equation}
where $U$ is an $N\times N$ unitary matrix \cite{GG, Ha}. The parameter $\ell>0$ fixes the length scale; keeping it fixed, we denote the Hamiltonian determined by the condition \eqref{bc} by $H_U$.

As usual in quantum mechanics, a symmetry is described by an operator $\HH\to\HH$ with respect to which the Hamiltonian is invariant. Since the symbol of our operator is real and invariant with respect to the orientation of the edges, the nontrivial part of the task concerns the condition \eqref{bc}: a particular symmetry is associated with an invertible map in the space of the boundary values, $\Theta:\,\C^n\to\C^n$, such that $(U-I)\Theta\Psi(0)+i(U+I)\Theta\Psi'(0)=0$ holds for all $\Psi$ satisfying \eqref{bc}, which is equivalent to the fact that $U$ obeys the identity
\begin{equation}\label{sym}
\Theta^{-1}U\Theta=U.
\end{equation}
The question is now which operators can be associated with the parity and time reversal transformations.

The latter is simpler. The operator $\Theta_\mathcal{T}$ is antilinear and idempotent, and since the particles we consider have no internal degrees of freedom such as spin, it is just the complex conjugation. Using the unitarity, $U^T\bar{U}=\bar{U}U^T=I$ we find easily that $\bar\Psi$ satisfies the condition \eqref{bc} with the transposed matrix, that is,
\begin{equation}\label{Trevers}
\Theta_\mathcal{T}^{-1}U\Theta_\mathcal{T}= \Theta_\mathcal{T}U\Theta_\mathcal{T}=U^T,
\end{equation}
and consequently, the $H_U$ is $\mathcal{T}$-invariant if and only the matrix defining the coupling is transposal-symmetric, $U=U^T$. This also immediately implies that a (self-adjoint) quantum graph \emph{is $\mathcal{PT}$-symmetric if and only if the mirror transformation acts in the same way,}
\begin{equation}\label{mirror}
\Theta_\mathcal{P}^{-1}U\Theta_\mathcal{P}=\Theta_\mathcal{P}U\Theta_\mathcal{P}=U^T.
\end{equation}

Asking under which circumstances can the condition \eqref{mirror} be satisfied, we have to note first that -- while the concept of quantum graph \emph{per se} does not need an ambient space -- investigation of spatial reflections forces us to think of the graph as embedded in the Euclidean space. For the sake of simplicity we regard our star graph as planar, but the conclusion will certainly extend to more general situations.

Let us first indicate which operators cannot serve as $\Theta_\mathcal{P}$. One might be tempted to reverse the edge orientation leading to flipping the sign of $\Psi'$, however, the edges are all parametrized in the same outward direction, which does not change when viewed in a mirror; moreover, such a change would turn, say, an attractive $\delta$ coupling to a repulsive one. Having a planar star graph, one can also think of reversing the edge numeration (as considered in \cite{ETT}), but the corresponding candidate for the role of $\Theta_\mathcal{P}$, the matrix with one on the main antidiagonal and zeros elsewhere, leads to a double transpose of $U$, both with respect to the diagonal and antidiagonal, meaning just renaming the edges.

To see which linear operator can facilitate the similarity between $U$ and $U^T$, we use the unitarity of the matrix: there is a unitary $V$ such that $VUV^*$ is diagonal, and as such equal to its transpose. It follows that the matrix $\Theta$ satisfying $\Theta U\Theta=U^T$ is of the form $\Theta = V^TV$. We know, however, how $V$ looks like: the $j$th column of $V^*$ coincides with $\phi_j^T$, where $\phi_j$ is the $j$th normalized eigenvector of $U$. Consequently, we have
\begin{equation}\label{Theta}
\Theta_{ij} = (\bar\phi_i, \phi_j), \quad i,j=1,\ldots,n\,;
\end{equation}
note the complex conjugation in the left entry which makes these expressions nontrivial. Denoting by $\{\nu_j\}$ the basis in the boundary value space referring to the direct sum form of $\HH$, $\nu_1=(1,0,\ldots,0)^T$ etc., we see that $\Theta$ maps $\nu_j$ to $((\bar\phi_1, \phi_j), \ldots (\bar\phi_n, \phi_j))^T$, so it general one cannot expect that such a $\Theta$ to be associated with a mirror transformation.

The situation changes, however, when we restrict our attention to vertex couplings referring to the subset of \emph{circulant} matrices. They are of the form
\begin{equation}\label{C}
U = \left(\begin{array}{ccccc}
c_{1} & c_{2} & \cdots & c_{n-1} & c_{n} \\
c_{n} & c_{1} & c_{2} &  & c_{n-1} \\
\vdots & c_{n} & c_{1} & \ddots & \vdots \\
c_{3} &  & \ddots & \ddots & c_{2} \\
c_{2} & c_{3} & \cdots & c_{n} & c_{1}
\end{array}\right)\,;
\end{equation}
the unitarity requires that
\begin{equation}\label{invDFT}
c_j=\frac{1}{n}\left(\lambda_1+\lambda_2\omega^{-j}+\lambda_3\omega^{-2j}+\cdots+\lambda_n\omega^{-(n-1)j}\right), \quad j=1,\ldots,n,
\end{equation}
where $\lambda_j,\:j=1,\ldots,n$, are the eigenvalues of $U$ and $\omega:=\mathrm{e}^{2\pi i/n}$. The corresponding normalized eigenvectors are independent of the choice of the $c_j$'s being
\begin{equation} \label{eigenv}
\phi_j=\frac{1}{\sqrt{n}}\left(1,\omega^j,\omega^{2j},\ldots,\omega^{(n-1)j}\right)^T, \quad j=1,\ldots,n.
\end{equation}
Furthermore, the eigenvalues can be written in terms of the matrix entries as $\lambda_j = \sum_{k=1}^n c_k \omega^{j(k-1)}$. The diagonalization is achieved in this case by the discrete Fourier transformation,
\begin{equation*}
V^* = \frac{1}{\sqrt{n}}\left(\begin{array}{cccccc}
1 & 1 & 1 & 1 & \ldots & 1 \\
1 & \omega & \omega^2 & \omega^3 & \ldots & \omega^{(n-1)} \\
1 & \omega^2 & \omega^4 & \omega^6 & \ldots & \omega^{2(n-1)} \\
\vdots & \vdots & \vdots & \vdots &  & \vdots \\
1 & \omega^{n-1} & \omega^{2(n-1)} & \omega^{3(n-1)} & \ldots & \omega^{(n-1)^2} \\
\end{array}\right)\,.
\end{equation*}
Using relations \eqref{Theta} and \eqref{eigenv} we easily find that now we have
\begin{equation}\label{Thetamirr}
\Theta_\mathcal{P} = \left(\begin{array}{ccccccc}
1 & 0 & 0 & \cdots & 0 & 0 & 0 \\
0 & 0 & 0 & \cdots & 0 & 0 & 1 \\
0 & 0 & 0 & \cdots & 0 & 1 & 0 \\
\vdots & & & \iddots &  & & \vdots \\
0 & 0 & 1 & \cdots & 0 & 0 & 0 \\
0 & 1 & 0 & \cdots & 0 & 0 & 0
\end{array}\right)\,;
\end{equation}
Now we may add the subscript $\mathcal{P}$ to $\Theta$ because the transformation has the properties expected from the parity transformation. It preserves the edge $e_1$, as well as $e_{k+1}$ if $n=2k$, and among the remaining ones it switches $e_j$ with $e_{n+2-j}$, and moreover, the same will be true if we renumber the edges.

One may ask what is the meaning of the vertex couplings characterized by matrices of the form \eqref{C}. It is clear that the corresponding operator $H_U$ is invariant with respect to a cyclic renumbering of the edges meeting at the vertex. This becomes illustrative if we think of our graph as embedded in the Euclidean space, for simplicity again as a planar star. If the angles between the edges are the same, each such renumbering is equivalent to a discrete rotations; in this sense we can regard such a system as \emph{isotropic}.

In this way we have identified a class of vertex couplings for which the graph exhibits a $\mathcal{PT}$-symmetry. It depends on $n$ real parameters, out of the number $n^2$ which characterize an arbitrary self-adjoint coupling. Among them, a subset depending on $\big[\frac{n}{2}\big]+1$ parameters is separately symmetric with respect to the time inversion and mirror transformation, while in the $\big[\frac{n-1}{2}\big]$-parameter complement the $\mathcal{PT}$-symmetry is nontrivial.

The former contains some well-known coupling such as $\delta$ or $\delta'_s$, or more generally, all the permutation-invariant ones which correspond to $U$ of the form $U=uI+vJ$, where $I$ is the unit matrix and $J$ denotes the matrix with all the entries equal to one; the parameters have to fulfil the conditions $|u|=1$ and $|u+nv|=1$. Among vertex couplings which violate the time-reversal symmetry, a simple example was proposed in \cite{ETa} being inspired by an attempt to use quantum graphs to model the anomalous Hall effect \cite{SK}. The indicated coupling is determined by the matrix
\begin{equation}\label{matU}
  U= R := \begin{pmatrix}0 & 1 & 0 & 0 & \cdots & 0 & 0\\ 0 & 0 & 1 & 0 & \cdots & 0 & 0\\ 0 & 0 & 0 & 1 & \cdots & 0 & 0\\ \vdots & \vdots & \vdots & \vdots & \ddots & \vdots & \vdots\\ 0 & 0 & 0 & 0 & \cdots & 0 & 1\\ 1 & 0 & 0 & 0 & \cdots & 0 & 0\\ \end{pmatrix},
\end{equation}
which is obviously circulant and transposition-asymmetric. The same is true for a one-parameter class of couplings introduced in \cite{ETT} which interpolates between the $\delta$ coupling, to which the matrix $U=-I+\frac{2}{n+i\alpha}J$ corresponds making it separately $\mathcal{P}$- and $\mathcal{T}$-symmetric, and the one referring to \eqref{matU}.

Spectral properties of a self-adjoint star graph do not depend on the particular form of the matrix $U$, only on its spectrum. The essential spectrum of $H_U$ is always the halfline $\R^+$, being absolutely continuous, because any of these Hamiltonians has a common symmetric restriction with $H_\mathrm{D}:= H_{-I}$ describing the graph with Dirichlet-decoupled edges. Its deficiency indices are at most $n$; this also means that the number of negative eigenvalues of $H_U$ does not exceed $n\:$ \cite[Corr. to Thm~8.19]{We}. In fact, we can find the eigenvalues explicitly using the fact that they are preserved by unitary equivalence. Choosing the orthonormal basis in $\HH$ referring to the eigenvectors of $U$, we get a decomposition of $H_U$ into a direct sum, the components of which are operators $-\frac{\D^2}{\D x^2}$ on the halfline with the boundary condition $(\e^{i\gamma_j}-1)\phi_j(0) + i\ell(\e^{i\gamma_j}+1)\phi'_j(0)=0$, where $\lambda_j= \e^{i\gamma_j},\: j=1,\ldots,n$, with $\gamma_j\in[0,2\pi)$ are the eigenvalues of $U$. Using the Ansatz $\phi_j(x)= \e^{-\kappa_j x}$ we see that $\#\sigma_\mathrm{disc}(H_U)$ \emph{coincides with the number of eigenvalues of $U$ is the upper complex halfplane} and they are of the form $-\kappa_j^2$, where
\begin{equation}\label{ev's}
\kappa_j = \frac{1}{\ell}\, \tan\frac{\gamma_j}{2}, \quad \gamma_j\in(0,\pi).
\end{equation}
Likewise, the eigenvalues of $U$ in the lower halfplane give rise to antibound states.

On the other hand, transport properties of such a star graph depend on the form of $U$, not just on its spectrum. In \cite{ETa} an interesting topological property of the coupling \eqref{matU} was observed: in the high-energy asymptotic regime the transport is nontrivial if the vertex degree is even, while for an odd degree the vertex becomes effectively decoupled. The origin of this property is clear from the eigenvalue structure of \eqref{matU}: in the notation of  \eqref{invDFT} its discrete spectrum consists of the numbers $\omega^j:=\mathrm{e}^{2\pi ij/n},\: j=0,\ldots,n-1$. In particular, $-1\not\in \sigma(U)$ for $n$ odd, in which case the on-shell S-matrix,
\begin{equation}\label{S(k)}
S(k)= \frac{(k\ell-1)I+(k\ell+1)U}{(k\ell+1)I+(k\ell-1)U},
\end{equation}
satisfies $\lim_{k\to\infty} S(k)= I$; in the even case $-1$ is an eigenvalue and the limit looks differently on the corresponding eigenspace. This difference has various important consequences, in particular, for the band and gap structure of periodic lattices and chains \cite{BET, ETa}, bulk-edge differences in transport on lattices with a boundary \cite{EL20}, as well as for high-energy properties of finite graphs spectra \cite{EL19}.

This example illustrates the importance of the decomposition of the vertex coupling into the Dirichlet (the eigen\-space referring to eigenvalue -1), Neumann (the eigenvalue 1 eigenspace) and Robin (the rest) parts \cite[Thm.~1.4.4]{BK}. One can find other examples involving circulant matrices. A very simple one refers to the replacement of \eqref{matU} with $U=-R$ in case of $n=3,\,\ell=1$. Introducing $\eta:= \frac{1-k}{1+k}$, we have $S(k)= \frac{-\eta I+U}{I-\eta U}$ which is easily calculated to be
\begin{equation}\label{S3}
  S(k) = \frac{1}{1-\eta+\eta^2} \begin{pmatrix} -\eta & \eta-1 & \eta(1-\eta) \\ \eta(1-\eta) & -\eta & \eta-1 \\ \eta-1 & \eta(1-\eta) & -\eta \end{pmatrix}.
\end{equation}
This time the Dirichlet component is present but the Neumann one is not, and consequently, we have $\lim_{k\to 0} S(k)=I$, while the transport at high energies is nontrivial,
\begin{equation}\label{S3lim}
  \lim_{k\to\infty} S(k) = \frac13 \left(\begin{array}{rrr} 1 & -2 & -2 \\ -2 & 1 & -2 \\ -2 & -2 & 1 \end{array}\right),
\end{equation}
differing just by the sign of the off-diagonal elements from the Kirchhoff coupling S-matrix.

On the other hand, the vertex coupling can have a purely Robin character. To illustrate this claim, we replace the matrix \eqref{matU} with
\begin{equation}\label{Ueps}
  U = \epsilon R, \quad \epsilon = \e^{i\mu}, \quad \mu\in\big(0,\textstyle{\frac{2\pi}{n}}\big)\,;
\end{equation}
in components the matching condition then reads
\begin{equation}\label{Ucomp}
  \epsilon\psi_{j+1}-\psi_j + i\ell(\epsilon\psi'_{j+1}+\psi'_j) = 0 \quad (\mathrm{mod}\,n)
\end{equation}
and its $\mathcal{PT}$-symmetry is obvious. Putting $\eta:= \frac{1-k\ell}{1+k\ell}$ we find for the S-matrix elements the expression
\begin{equation}\label{Selem}
  S_{ij}(k) = \frac{1}{1-\epsilon^n\eta^n} \Big( -\eta(1-\epsilon^n\eta^{n-2})\delta_{ij} + (1-\delta_{ij}) (1-\eta^2)\epsilon (\epsilon\eta)^{(j-i-1)(\mathrm{mod}\,n)} \Big),
\end{equation}
which reduces for $\epsilon=1$ to the appropriate formula from \cite{ETa}. In distinction to that case, however, we have now $\lim_{k\to\infty} S(k)= I$ because of the factor $1-\eta^2$ which cancels out with the prefactor only if $\epsilon=1$.

To see how the presence of the phase factor in \eqref{Ueps} influences the spectrum, let us revisit an example discussed in \cite{ETa}, a periodic \emph{square lattice} of edge length $\ell$. To make the comparison easier, we put the length-scale parameter in \eqref{Ucomp} equal to one, which we can do without loss of generality, because such a choice is by the natural scaling transformation equivalent to keeping the $\ell$ and setting the lattice spacing to one. The way to find the spectrum is the same as in the said paper: we use the Ansatz
\begin{align}
\psi_1(x) &= a_1\e^{ikx} + b_1\e^{-ikx}, \nonumber \\
\psi_2(x) &= a_2\e^{ikx} + b_2\e^{-ikx}, \nonumber  \\[-.7em] \label{Ansatz} \\[-.7em]
\psi_3(x) &= \omega_1 \left( a_1\e^{ik(x+\ell)} + b_1\e^{-ik(x+\ell)} \right), \nonumber \\
\psi_4(x) &= \omega_2 \left( a_2\e^{ik(x+\ell)} + b_2\e^{-ik(x+\ell)} \right), \nonumber
\end{align}
for wave functions in the elementary cell of the square lattice having the cross-shaped form with four edges of length $\frac12\ell$. We substitute the corresponding boundary values into \eqref{bc} remembering that the derivatives have to be taken in the outward direction, and require the values at the `loose' ends of the cross to differ by the Bloch factors $\omega_j=\e^{i\theta_j}$, where $\theta_j,\, j=1,2$, are the quasimomentum components. This yields a system of four linear equations for the coefficients $a_j,\,b_j$ which is solvable provided the determinant
\begin{equation} \label{determinant}
D \equiv D(\eta,\omega_1,\omega_2) = \left|
\begin{array}{cccc}
-1 & -\eta & \epsilon\eta & \epsilon \\[.2em]
\epsilon\omega_1\xi^2 & \epsilon\omega_1\bar\xi^2\eta & -1 & -\eta
\\[.2em]
-\omega_1\xi^2\eta & -\omega_1\bar\xi^2 & \epsilon\omega_2\xi^2 & \epsilon\omega_2\bar\xi^2\eta \\[.2em]
\epsilon\eta & \epsilon & -\omega_2\xi^2\eta & -\omega_2\bar\xi^2
\end{array}
\right|
\end{equation}
with $\xi=\e^{ik\ell}$ and $\epsilon = \e^{i\mu}$ with $\mu\in\big(0,\frac12\pi\big)$ vanishes. Passing from $\eta$ to the original momentum variable $k$ and using the trigonometric expressions for $\epsilon^4\pm 1$ we arrive at the condition
\begin{equation} \label{detcond}
8i\epsilon^2\,\frac{\e^{i(\theta_1+\theta_2)}}{(k+1)^4}\, \sum_{j=0}^4 c_jk^j = 0,
\end{equation}
where
\begin{align}
c_0 &= c_4 = -\sin 2\mu\,\sin^2 k\ell, \nonumber \\
c_2 &= \sin 2\mu\, (1+3\cos 2k\ell), \nonumber  \\[-.7em] \label{coeff} \\[-.7em]
c_1 &= 2\big( 2\cos 2\mu\, \cos k\ell -\cos\theta_1 - \cos\theta_2\big) \sin k\ell, \nonumber \\
c_3 &= 2\big( 2\cos 2\mu\, \cos k\ell +\cos\theta_1 + \cos\theta_2\big) \sin k\ell; \nonumber
\end{align}
for the negative spectrum we set $k=i\kappa$ with $\kappa>0$ and the spectral condition becomes $\sum_{j=0}^4 c_j\kappa^j =0$, where
\begin{align}
c_0 &= c_4 = \sin 2\mu\,\sinh^2 \kappa\ell, \nonumber \\
c_2 &= -\sin 2\mu\, (1+3\cosh 2\kappa\ell), \nonumber  \\[-.7em] \label{coeff-} \\[-.7em]
c_1 &= -2\big( 2\cos 2\mu\, \cosh \kappa\ell -\cos\theta_1 - \cos\theta_2\big) \sinh \kappa\ell, \nonumber \\
c_3 &= 2\big( 2\cos 2\mu\, \cosh \kappa\ell +\cos\theta_1 + \cos\theta_2\big) \sinh \kappa\ell; \nonumber
\end{align}
This shows that the presence of the phase factor influences the spectrum significantly. If $\mu=0$ the even coefficients are zero and the corresponding solution to \eqref{detcond} given in \cite{ETa} shows that the positive spectrum is dominated by spectral bands the widths of which grow linearly with band index, while the gap widths are asymptotically constant. Moreover, the factor $\sin k\ell$ in $c_1$ and $c_3$ means that the spectrum contains
infinitely degenerate eigenvalues, or flat bands.

This changes once we have $\mu\ne 0$, within the indicated interval. First of all, the form of $c_2$ does not allow to factorize a $\theta$-independent term from \eqref{detcond} which means that \emph{there is no infinite series of flat bands} such as the one at $k^2=\big(\frac{\pi n}{\ell}\big)^2,\, n\in\Z$, appearing for $\mu=0$. This does not mean that the point spectrum is void, though. Choosing $k=1$, the spectral condition reduces to $\sum_{j=0}^4 c_j = 0$ where the $\theta$-dependent terms cancel and the sum vanishes provided
\begin{equation} \label{detcond}
\cot2\mu = \frac{-1 - 3\cos2\ell +2\sin^2\!\ell}{4\sin2\ell} = - \cot2\ell,
\end{equation}
that is, $\mu=\frac{\pi}{2}-\ell\; \big(\mathrm{mod}\,\frac{\pi}{2}\big)$. Furthermore, while we lack now the nice graphical way in which we were able to solve the spectral problem in \cite{ETa}, it is not difficult to determine the high-energy asymptotic behavior. Since $c_4\ne 0$ for $k\ne\frac{\pi n}{\ell}$, it is clear from \eqref{detcond} that -- while $\big(\frac{\pi n}{\ell}\big)^2\not\in\sigma(H_U)$ -- the spectral bands may exist only in the vicinity of those points. Up to higher-order terms, $k=\frac{\pi n}{\ell}+\delta$ may correspond to a spectral point only for $\delta$ within the range of the function $(\theta_1,\theta_2) \mapsto \frac{\pi n(2\cos2\mu+(-1)^nQ)}{(\pi^2n^2 +6\ell^2)\sin2\mu}$, where $Q:=\cos\theta_1+\cos\theta_2$ runs through the interval $[-2,2]$. For large $n$ we thus get $|\delta_n|\lesssim \frac{2}{\pi n}\cot\mu$ as $n\to\infty$, and consequently, the width of the $n$th band on the energy scale is \emph{for a fixed} $\mu\in\big(0,\textstyle{\frac{\pi}{2}}\big)$ asymptotically constant,
\begin{equation} \label{bandwidth}
\Delta_n \lesssim \frac{8}{\ell}\,\cot\mu.
\end{equation}
We stress the fixed value of $\mu$. The band width is decreasing as $\mu$ grows from zero, however, it is not monotonous over the whole interval $\big[0,\textstyle{\frac{\pi}{2}}\big]$; as we are approaching the right endpoint, it starts growing again, because the corresponding matrix $U=iR$ has again $-1$ as its eigenvalue which means that the spectrum is anew dominated by the bands.

This non-uniform character of the asymptotical behavior is also manifested in another quantity, the probability of finding a randomly chosen value of the energy in the spectrum,
\begin{equation}\label{prob}
P_{\sigma}(H_U):=\lim_{K\to\infty} \frac{1}{K}\left|\sigma(H_U)\cap[0,K]\right| = 0,
\end{equation}
introduced by Band and Berkolaiko \cite{BaB}. It is clear from \eqref{bandwidth} that $P_{\sigma}(H_U)=0$ for $\mu\in\big(0,\textstyle{\frac{\pi}{2}}\big)$, while for both the real-valued $U=R$ and the purely imaginary $U=iR$ the probability is equal to one.

One can also inspect the negative spectrum of $H_U$, which has at most two bands. In particular, for large $\ell$ they are expected to be narrow and to shrink to the eigenvalues \eqref{ev's} referring to $\kappa=\tan\frac{\mu}{2}$ and $\tan\big(\frac{\mu}{2} + \frac{\pi}{4}\big)$ as $\kappa\to\infty$. This is indeed the case: for large $\ell$ the leading behavior in the spectral condition with the coefficients \eqref{coeff-} is determined by the terms containing $\mathrm{e}^{2\kappa\ell}$, which yields
\begin{equation} \label{negasympt}
0 = \kappa^4+1 -6\kappa^2 +4\kappa(\kappa^2-1)\cot2\mu + \OO(\mathrm{e}^{-2\kappa\ell}),
\end{equation}
where only the error term depends on the quasimomentum. If we neglect it and look for solutions of the resulting equation, it us useful to put $c:=\cos\frac{\mu}{2}$ and $s:=\sin\frac{\mu}{2}$, so that $\cos2\mu= 1-8c^2s^2$ and $\sin2\mu= 4cs(c^2-s^2)$; by a straightforward computation one can then check that $\kappa=\frac{s}{c}$ and $\kappa=\frac{c+s}{c-s}$ are the roots.

\begin{figure}[htbp]
     \centering
     \includegraphics[clip, trim=0cm 0cm 0cm 0cm,width=0.5\textwidth]{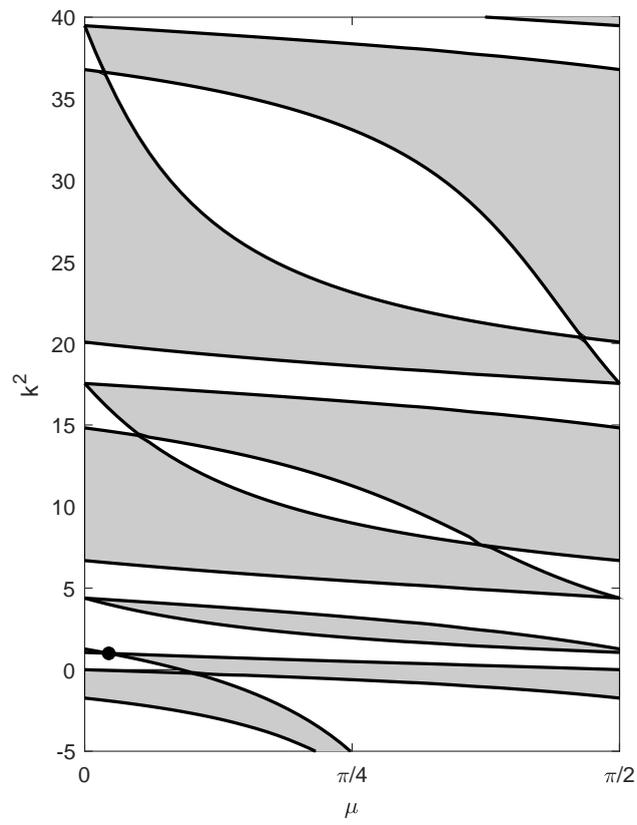}
     \caption{The spectrum of $H_U$ as a function of $\mu$ for $\ell=\frac32$. The dot at $(\mu,k)=\big(\frac12(\pi-3),1\big)$ is the flat band.}\label{fig32}
\end{figure}
\begin{figure}[htbp]
     \centering
     \includegraphics[clip, trim=0cm 0cm 0cm 0cm,width=0.5\textwidth]{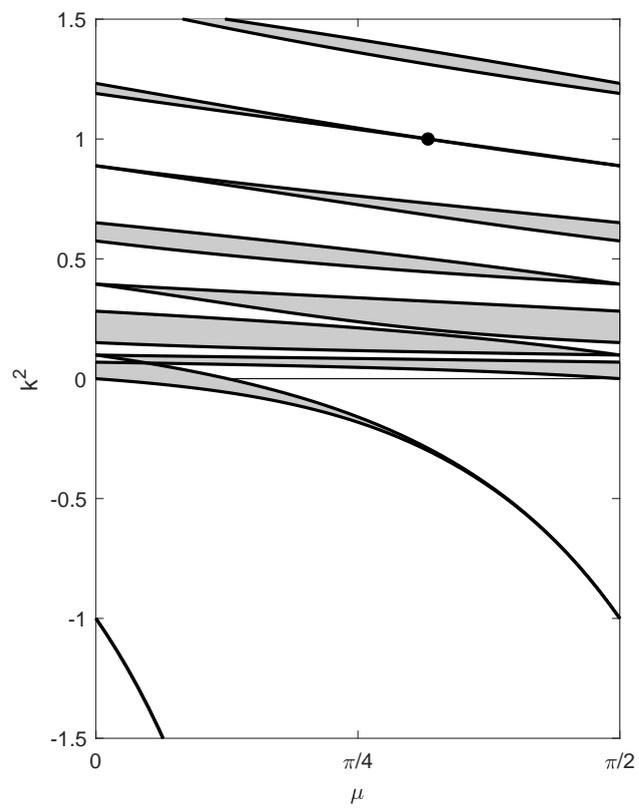}
    \caption{The same for $\ell=10$.}\label{10low}
\end{figure}
\begin{figure}[htbp]
     \centering
     \includegraphics[clip, trim=0cm 0cm 0cm 0cm,width=0.5\textwidth]{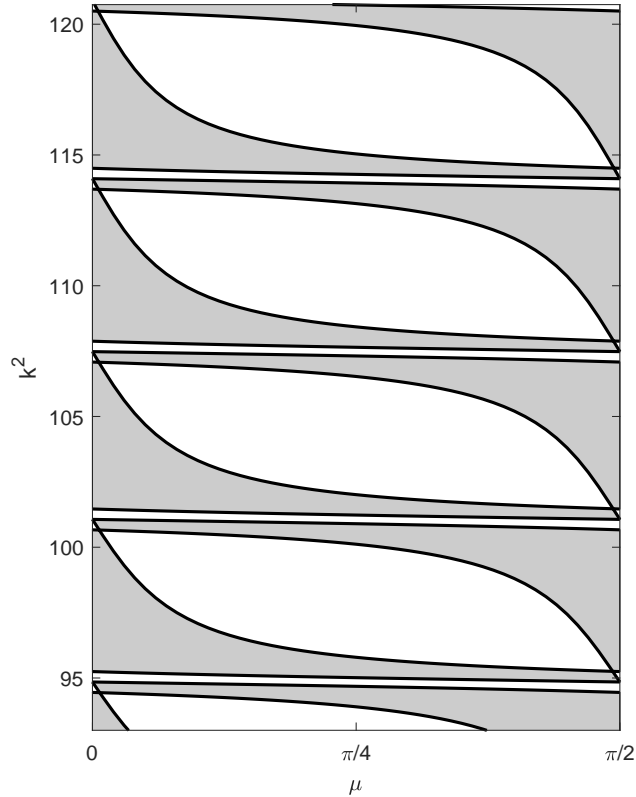}
     \caption{High energy spectrum of $H_U$ as a function of $\mu$ for $\ell=10$.}\label{10high}
\end{figure}
\begin{figure}[htbp]
     \centering
     \includegraphics[clip, trim=0cm 0cm 0cm 0cm,width=.8\textwidth]{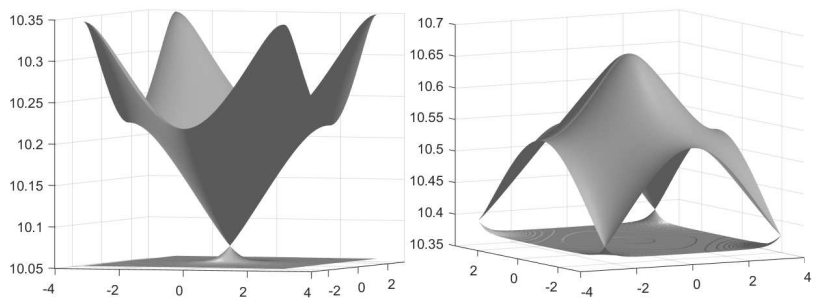}
     \caption{Fermi surfaces in the momentum variable for the lattice with $\ell=10$ at the points of closing gaps, left at the values $(\mu,k)=(1.55068665,10.07328547)$, right at $(\mu,k)=(1.55190524,10.38681556)$}\label{Dcone}
\end{figure}

These results are illustrated in Fig.~\ref{fig32} in which we plot the spectrum of $H_U$ in dependence of the parameter $\mu$ with the choice $\ell=\frac32$, including the indication of the flat band. Fig.~\ref{10low} shows the same spectrum for a larger value of the length parameter, $\ell=10$. This choice makes the bands narrow, in particular, it is clear that the negative ones are centered around the corresponding star-graph eigenvalues \eqref{ev's} obtained as solutions to \eqref{negasympt}. The flat band at $(\mu,k)=(0,-1)$ becomes absolutely continuous for $\mu>0$ and escapes to negative infinity as $\mu\to\frac{\pi}{2}-$ while the next one shrinks to $-1$ in this limit; recall that at that point the matrix $U$ has eigenvalue $-1$ again.
In Fig.~\ref{10high} we plot a higher part of the same spectrum. The picture illustrates that for a fixed $\mu\in\big(0,\frac{\pi}{2}\big)$ the positive spectral bands get narrower as the energy grows, while at the endpoints of the interval they dominate the spectrum; as expected the limit \eqref{prob} is nonuniform with respect to $\mu$.
In addition, we see a number of points where the gaps are closing. In distinction to the point $(\mu,k)= \big(\frac{\pi}{2}-\ell,1)$ mentioned above, however, those are not flat bands but true band crossings occurring either in the center of the Brillouin zone or its corners. Two examples are shown in Fig.~\eqref{Dcone}; one clearly sees the Dirac cones at the touching points.

\section*{Acknowledgements}
The research was supported by the Czech Science Foundation within the project 21-07129S and by the EU project CZ.02.1.01/0.0/0.0/16\textunderscore 019/0000778. 


\end{document}